\documentclass[english]{iopart}
\usepackage[T1]{fontenc}
\usepackage[latin9]{inputenc}
\usepackage{textcomp}
\usepackage{amsbsy}
\usepackage{amstext}
\usepackage{amssymb}
\usepackage{graphicx}

\makeatletter
\usepackage{iopams}
\usepackage{setstack}

\makeatother

\usepackage{babel}
\begin{document}

\title{Searching for New Physics Through AMO Precision Measurements}

\author{Chad Orzel}

\address{Union College Department of Physics and Astronomy, Science and Engineering
Center, Schenectady, NY 12308 USA}

\ead{orzelc@union.edu}
\begin{abstract}
We briefly review recent experiments in atomic, molecular, and optical
physics using precision measurements to search for physics beyond
the Standard Model. We consider three main categories of experiments:
searches for changes in fundamental constants, measurements of the
anomalous magnetic moment of the electron, and searches for an electric
dipole moment of the electron.
\end{abstract}
\maketitle

\section{Introduction}

Most physicists asked to think of searches for physics beyond the
Standard Model immediately picture multi-billion dollar particle accelerators
with detectors the size of office buildings, or gigantic astrophysical
detectors in remote locations. This is a natural assumption, as very
high energies are required to produce and detect exotic particles
associated with new physics. At lower energies, interactions with
these particles are very weak, and thus they have little effect on
physics at typical atomic energy scales of a few electron volts.

In fact, though, in labs around the world, new physics searches are
underway using atoms and molecules. These experiments, to borrow a
phrase from Gerald Gabrielse of Harvard, search for new physics using
precision, rather than energy. While the effects of new physics at
atomic energy scales are tiny, modern laser spectroscopy allows measurements
of astonishing precision, sufficient to detect the subtle influence
of new fundamental physics.

In this paper, we give a brief review of the essential techniques
involved in AMO precision measurements, then look at three types of
new-physics searches through precision measurement: the use of atomic
clocks to look for changes in fundamental constants, measurements
of the anomalous magnetic moment of the electron, and searches for
a permanent electric dipole moment of the electron.

\section{Precision Spectroscopy}

The general problem of precision spectroscopy is to find the frequency
of light needed to drive the transition between two quantum states.
For extremely precise measurements, these must be states with a small
natural linewidth due to spontaneous emission; pairs of atomic hyperfine
states, or dipole-forbidden transitions are typical. In addition to
the two states being studied, there must be a state-selective detection
method, typically involving laser excitation via a dipole-allowed
transition to another state and detection of the resulting fluorescence.

In the last several decades, methods of precision spectroscopy have
advanced to the point where transition frequencies can be determined
at the level of a few parts in $10^{18}$, sufficient to resolve possible
contributions from physics beyond the Standard Model. This level of
precision relies on two principal tools: the Ramsey separated-fields
method of spectroscopy, and the femtosecond laser frequency comb.
In this section, we will briefly review these tools, before moving
on to discuss the experiments that apply them.

\subsection{Ramsey Interferometry\label{sub:Ramsey-Interferometry}}

Ultimately, AMO precision measurement techniques can be traced to
the development of highly accurate atomic clocks. The laser-cooled
cesium clocks that are presently used as primary time standards are
stable to within a few parts in $10^{16}$\cite{CsFountains} and
improvements of an order of magnitude or more have been demonstrated
with experimental clocks based on other atoms. All of these clocks
are based on the separated-field spectroscopy method developed by
Norman Ramsey in 1950\cite{Ramsey}, and most AMO precision measurements
use some version of the Ramsey technique, so we briefly review it
here.

In Ramsey interferometry, a system in state $|1>$ is first exposed
to light at frequency $\omega$ near the resonant frequency $\omega_{0}$,
which drives coherent Rabi oscillations between $|1>$ and $|2>$
at a frequency $\Omega.$ The intensity and duration of the pulse
are chosen to prepare the system in a coherent superposition of $|1>$
and $|2>$(a ``$\pi/2$-pulse''). The superposition evolves freely
for a time$T$, then is exposed to a second $\pi/2$ pulse. At the
end of this pulse sequence, the state of the system is probed to determine
whether it has made a transition to $|2>$. 

The transition probability at the end of this sequence is

\[
P(1\rightarrow2)=\left(\frac{\Omega\tau_{\pi/2}}{2}\right)^{2}\left[\frac{\sin\frac{(\omega-\omega_{0})}{2}\tau_{\pi/2}}{\frac{(\omega-\omega_{0})}{2}\tau_{\pi/2}}\right]^{2}\cos^{2}\left(\frac{(\omega-\omega_{0})}{2}T\right)
\]
 This consists of an overall amplitude depending on the Rabi frequency
$\Omega$ and the pulse duration $\tau_{\pi/2}$ and a rapid oscillation
depending on the free evolution time $T$. The width in frequency
of one of these ``Ramsey fringes'' is $\Delta\omega=\frac{\pi}{T}$,
inversely proportional to the free evolution time $T$. This is what
gives the Ramsey method its extreme sensitivity: for laser-cooled
Cs clocks, $T\sim1\,$s\cite{CsFountains}, allowing frequency resolution
of less than $1$Hz on the $9.19$GHz transition between Cs hyperfine
ground states that defines the SI second.

\begin{figure}
\includegraphics[scale=0.4]{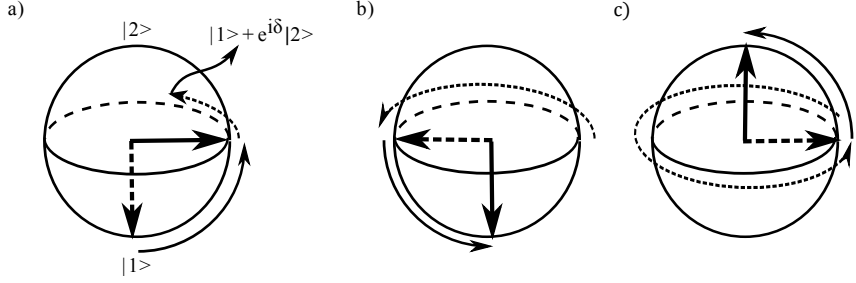}\caption{\label{fig:Ramsey}Bloch sphere visualization of the Ramsey spectroscopy
sequence. a) The initial $\pi/2$ pulse rotates the state vector into
a superposition state, which begins to precess about the axis at a
frequency $(\omega-\omega_{0})$. b) After a half-integer number of
rotations, the second $\pi/2$ pulse rotates the state vector back
to state $|1>$. c) After an integer number of rotations, the second
$\pi/2$ pulse completes the transition to state $|2>$. }
\end{figure}

The Ramsey fringes arise from interference between the two parts of
the initial superposition, whose phases evolve at different rates.
An intuitive illustration uses the Bloch sphere picture (Fig.\ref{fig:Ramsey}),
in which the state is represented by a vector to a point on the surface
of a unit sphere. The states $|1>$and $|2>$correspond to the south
and north poles of the sphere, with other surface points describing
superposition states. A $\pi/2$ pulse corresponds to rotating the
state vector by $\pi/2$ about the x-axis, so the initial pulse of
the Ramsey sequence rotates the vector from the south pole to the
equator of the Bloch sphere, along the +y-axis. If this were followed
immediately by the second $\pi/2$ pulse, the state vector would be
rotated another $\pi/2$, ending at the north pole (the $|2>$state).

During the free evolution time, however, the phase of the superposition
evolves at a frequency $\omega-\omega_{0}$ (in the rotating wave
approximation), corresponding to precession of the state vector about
the z-axis. The result of the second $\pi/2$ pulse thus depends on
the position after the free evolution time $T.$ If the state vector
has completed an integer number of rotations, it returns to its initial
position, and the second $\pi/2$ rotation completes the transition
from $|g>\rightarrow|e>$. If the state vector has completed a half-integer
number of rotations, however, the state vector will be along the -y-axis,
and the second $\pi/2$ rotation returns it to the south pole, returning
the system to the ground state.

The oscillating coherence between the terms of the superposition state,
then, is the origin of the $T$-dependence of the Ramsey fringes.
Most AMO precision measurements use some variant of the Ramsey scheme:
an initial interaction to prepare a coherent superposition between
two states of an atomic or molecular system, followed by a period
of free evolution, followed by a second interaction with the light
field, then a state measurement, with high frequency resolution enabled
by interference between states.

\subsection{Optical Frequency Combs}

As the SI definition of time is based on the hyperfine ground state
splitting in Cs, measurements of absolute frequencies must ultimately
be referenced to Cs clocks, but comparing transition frequencies in
the optical domain to the microwave Cs frequency has presented a significant
technical challenge. In recent years, the development of femtosecond
laser frequency combs has greatly simplified such comparisons, leading
to a rapid advance to measurements of unprecedented precision\cite{CombReview}.

\begin{figure}
\includegraphics[scale=0.4]{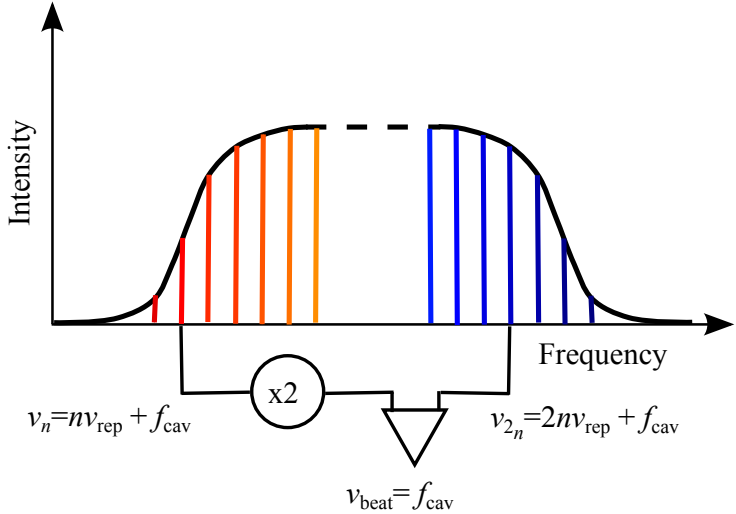}\caption{\label{fig:comb}A self-referenced frequency comb. Light from the
$n^{th}$ mode is frequency doubled and mixed with light from the
$2n^{th}$ mode. The beat frequency is equal to the offset frequency
due to cavity dispersion, allowing the absolute determination of any
optical frequency in terms of the cavity offset and the repetition
rate, both RF frequencies that are easily referenced to microwave
frequency standards.}
\end{figure}

The frequency spectrum of a mode-locked laser consists of a large
number of regularly spaced narrow modes (Fig.\ref{fig:comb}) , The
$n^{th}$ mode has a frequency

\[
\nu_{n}=n\nu_{rep}+f_{cav}
\]
where $\nu_{rep}$ is the repetition frequency of the laser (typically$\sim100\,$MHz),
which is controlled by the laser cavity, and $f_{cav}$ is a frequency
offset due dispersion in the cavity. The width of the frequency spectrum
will be determined by the duration of a single laser pulse, and for
pulses a few femtoseconds long, the bandwidth of the comb can span
a full octave of frequency. Light from the $n^{th}$ mode can be frequency-doubled
and mixed with light from the $2n^{th}$ mode, producing a radio-frequency
beat note at the difference frequency:

\[
\Delta\nu=2\nu_{n}-\nu_{2n}=(2n\nu_{rep}+2f_{cav})-(2n\nu_{rep}+f_{cav})=f_{cav}
\]
This comparison directly measures the cavity offset, allowing the
absolute determination of the frequency of any mode in terms of RF
frequencies that can easily be referenced to a microwave atomic clock.
Such a self-referenced frequency comb allows the direct comparison
of two different laser frequencies across a wide range: the repetition
rate of the comb can be locked to an atomic clock, and the absolute
laser frequencies determined from the beat note with the nearest tooth
of the comb. Aternatively, the comb can be stabilized to one of the
two laser sources (from an optical-frequency atomic clock, for example),
and the relative frequency difference determined from the beat note
of the second laser.

Frequency combs have found applications in molecular spectroscopy\cite{CombSpectroscopy},
and as reference sources for astronomical spectrometers\cite{CombAstronomy,CombAstronomyPhillips}.
Their primary impact, however, has been in precision spectroscopy,
where the exceptional precision of comb-based comparisons allows the
possibility of frequency metrology to 18 or 19 decimal places, opening
many possibilities for new physics searches with AMO techniques.

\section{Changing Fundamental Constants}

Many theories of attempting combine gravity with the Standard Model
predict possible changes in the values of fundamental constants over
cosmological time scales\cite{UzanConstantsTheory}. These changes
are detectable through variation of dimensionless ratios of constants,
the most significant of which are the fine structure constant $\alpha=e^{2}/4\pi\epsilon_{0}\hbar c$,
which combines the electron charge $e$, speed of light $c$ and Planck's
constant $\hbar$, and the proton-electron mass ratio $\mu=m_{p}/m_{e}$.
Changes in these ratios change the relative energies of atomic and
nuclear states, reviewed in Ref. \cite{ChibaConstantsReview}; here
we will highlight only a few AMO-based searches. 

The influence of changing constants on nuclear states is seen primarily
through geophysical records of radioactive decay. Isotope ratios from
the ``natural nuclear reactor'' at Oklo provide a fossil record
of nuclear reaction rates during its operation 1.7 billion years ago,
limiting the possible variation to $-0.11\times10^{-7}\leq\Delta\alpha/\alpha\leq0.22\times10^{-7}$\cite{Lamoreaux}.
Iron meteorites containing records of the $\beta$-decay of $^{187}$Re
into $^{187}$Os weaker bound, $\Delta\alpha/\alpha=(2.5\pm16)\times10^{-7}$
over $4.6$ billion years\cite{meteorites}. Both of these methods
are somewhat model-dependent, as the reaction rates depend on the
environment in the distant past. Weaker constraints over cosmological
time scales are provided by analysis of the cosmic micrwave background\cite{WMAPalphadot,CMBAlphaDot}
and primordial nucleosynthesis\cite{NucleosynthAlphaDot}.

For atomic and molecular states, the fractional shift of a transition
frequency $\omega$ can be written in the general form (following
Ref. \cite{Lea}):

\begin{equation}
\frac{1}{\omega}\frac{\partial\omega}{\partial t}=A\frac{1}{\alpha}\frac{\partial\alpha}{\partial t}+B\frac{1}{\mu}\frac{\partial\mu}{\partial t}+C\frac{1}{g}\frac{\partial g}{\partial t}-\frac{1}{\omega_{Cs}}\frac{\partial\omega_{Cs}}{\partial t}\label{eq:freq_dep}
\end{equation}
where the numerical factors $A$,$B$, and $C$ give the sensitivity
of the transition in question to changes in $\alpha$, $\mu,$ and
the nuclear g-factor $g$, respectively. The latter two primarily
affect hyperfine transitions, which depend on the interaction between
the electron spin and the nuclear magnetic moment. These sensitivity
factors vary between different atoms and different transitions within
the same atom, and are determined from numerical calculations of atomic
or molecular wavefunctions. The final term accounts for the shift
in the frequency of the Cs ground-state hyperfine transition, which
provides the definition of the SI second, and is itself subject to
change with changes in the fundamental constants.

Spectroscopic observations of high-redshift astronomical objects such
as quasars and gas clouds provide a direct probe of the history of
the fine-structure constant. The general procedure is to compare the
observed wavelengths of spectral lines associated with a particular
astronomical object (either emission lines in the spectrum of the
object itself, or absorption features in the spectrum from a more
distant object such as a quasar) to determine whether the spectral
lines from an earlier epoch are the same as those for the same system
today. This procedure is complicated by the need to account for the
cosmological redshift due to the Hubble expansion, so these studies
generally look at differences between two or more transitions associated
with the same object.

The simplest spectroscopic measurements of the past history of fundamental
constants use pairs of lines associated with a single element; for
example, Cowie and Songaila\cite{Cowie} measured the difference between
fine structure states of Si IV, which scales as $\alpha^{2}$, obtaining
a value of $\Delta\alpha/\alpha=(3.3\pm5.5)\times10^{-6}$ at redshifts
of $z\sim1.8$ ($\sim10Gyr$). Numerous subsequent measurements have
had similar sensitivity\cite{Levshakov,Rahmani,Kanekar 2010,Kanekar 2012,Levshakov 2012},
at redshifts up to $z=5.2$, some finding significant changes, other
consistent with no change. 

The most dramatic of these results is the claim by Webb et al.\cite{Webb,King}
of not only a change over time, but a spatial variation having a dipole
pattern:

\[
\frac{\Delta\alpha}{\alpha}=Ar\cos\theta
\]
where $r=ct(z)$ is the look-back distance in Glyr, $\theta$is the
angle from the pole of the dipole pattern, and $A=(1.1\pm0.25)\times10^{-6}$Glyr$^{-1}$.
This claim is based on a many-multiplet analysis of data from the
Keck and VLT telescopes, using multiple spectral lines of several
different elements. Such a dipole pattern might explain the conflict
between measurements (depending on the location of the sources on
the sky), but it remains controversial\cite{Rahmani,Levshakov 2012,BayesianAlphaDot},
and awaits confirmation by independent observations or additional
telescopes.

\begin{figure}
\includegraphics[scale=0.4]{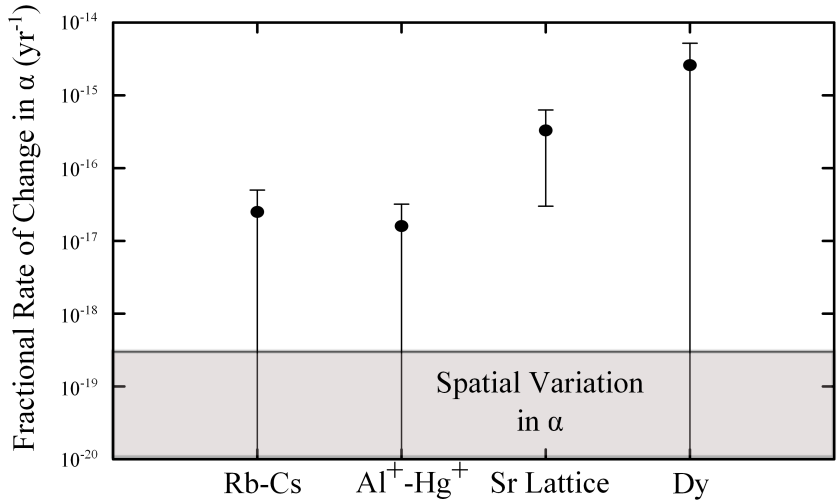}

\caption{\label{fig:AlphaDot}Summary of the best current measurements of $\dot{|\alpha}|/\alpha$
from AMO techniques. The horizontal line indicates the approximate
yearly rate of change due to the motion of the Sun through the spatial
gradient of Ref. \cite{Webb}. All the measurements are consistent
with $\dot{\alpha}=0$ at the $1-\sigma$ level except the Sr lattice
clock comparison from Ref. \cite{SrClockNetwork}.}
\end{figure}

Present-day spectroscopic measurements can not, of course, provide
any information about changes over cosmological time scales. They
do, however, provide a complementary measurement of the contemporary
rates of change, and the exceptional accuracy of modern atomic clocks
allows these present-day limits to be competitive with the best astrophysical
measurements. A measured frequency shift at the $1$ppm level looking
back $10^{10}$ years limits the average rate of change to $10^{-16}yr^{-1}$;
an atomic clock with accuracy of a part in $10^{16}$, like laser-cooled
Cs fountain clocks, needs only one year to match that limit. If the
dipole pattern claimed by Webb et al. exists, the motion of the solar
system through that background should produce a fractional change
of order $10^{-19}yr^{-1}$\cite{Berengut}, which may be within reach
of some AMO experiments (see Fig. \ref{fig:AlphaDot}) .

Reference \cite{Lea}provides an excellent summary of the history
of atomic-clock-based searches for changing constants as of 2007.
In this paper, we will focus on more recent measurements that demonstrate
the power of precision AMO techniques.

\subsection{Microwave Clock Comparisons}

Atomic cesium provides the basis of the SI second, defined as $9,192,631,770$
oscillations of the light associated with the ground-state hyperfine
transition in $^{133}$Cs, and laser-cooled fountain clocks now serve
as primary frequency standards for most national laboratories. The
analogous transition in rubidium, with a frequency of $6.8$GHz, has
also been studied as a possible time standard, due to some favorable
collisional properties. The LNE-SYRTE lab in Paris has been operating
an ensemble of fountains, including a dual-species apparatus capable
of operating with either Cs or Rb, since 1998. Measurements of the
frequency ratio $\eta=\nu_{Rb}/\nu_{Cs}$ over the last 14 years\cite{Marion,Guena}
provide one of the tightest constraints on modern changes $\alpha$.

As hyperfine transitions, the Cs and Rb transition frequencies are
sensitive not only to changes in $\alpha,$ but also to changes in
$g$ and $\mu$ as seen in equation \ref{eq:freq_dep}. The two transitions
have the same dependence on $\mu$, though, so the frequency ratio
is primarily sensitive to changes in $\alpha$ and weakly sensitive
to changes in the g-factors. The 14-year comparison finds a fractional
change$\dot{\eta}/\eta=(-1.36\pm0.91)\times10^{-16}yr^{-1}$. Combining
this result with the results of 6 other frequency ratio measurements,
which have different sensitivities to the various constants, allows
them to separate out the different components, giving $\dot{\alpha}/\alpha=(-0.25\pm0.26)\times10^{-16}yr^{-1}$
and $\dot{\mu}/\mu=(1.5\pm3.0)\times10^{-16}yr^{-1}$. These are comparable
to the best limits obtained from astrophysical measurements, and are
the second-best laboratory limit to date.

The very long observation time for the Rb/Cs comparison also allows
tests of other possible sources of variation. Binning their measurements
by month, they search for a possible variation with respect to the
position of the Earth in its orbit around the Sun, interpreted as
a possible coupling to the gravitational potential. They find a limit
of $c^{2}\frac{1}{\eta}\frac{\partial\eta}{\partial U}=(0.11\pm1.04)\times10^{-6}$,
an improvement by a factor of 1.4 over the best previous limit\cite{Ashby}.
The Rb-Cs comparison also limits any variation due to passage through
a spatial gradient in $\alpha$ to around $10^{-29}m^{-1}$, still
approximately two orders of magnitude larger than the expected variation
\cite{Berengut}. The Rb and Cs clocks are unlikely to improve by
this much, but other experiments may reach the necessary sensitivity.

\subsection{Optical Clocks}

One major area of time standards research over the last 30 years has
been the development of time standards operating at optical rather
than microwave frequencies. The move from GHz to THz frequencies can,
in principle, produce a dramatic increase in the precision of a clock,
but numerous technological developments needed to be made for optical
clocks to reach their full potential, such as laser sources with sub-Hz
linewidths, and optical frequency combs to connect the THz frequencies
of optical transitions to lower frequencies that can be compared to
current microwave time standards and more easily counted electronically
for use as a reference standard.

As these technologies have come together, numerous ``optical clock''
systems have been proposed and investigated, many of them using single
trapped ions\cite{IonClockReview} such as Hg$^{+}$\cite{HgClock},
Yb$^{+}$\cite{YbIonClock,YbIonOctupoleClock}, Sr$^{+}$\cite{SrIonClock},
In$^{+}$\cite{InIonClock} and Ca$^{+}$\cite{CaIonClock}. Trapped
ions can be laser cooled to the ground state of the trap, held for
long periods allowing long interaction times with a single ion, and
interrogated with high fidelity, making forbidden transitions to long-lived
metastable states a promising basis for an optical frequency standard.
Trapped-ion clocks in Yb$^{+}$\cite{PeikYbCs} and Hg$^{+}$\cite{FortierHgCs}
have been compared to microwave frequency standards in order to place
limits on changes in fundamental constants at the $\dot{\alpha}/\alpha\leq10^{-15}$
level.

The best performance of an optical clock to date is the $^{27}$Al$^{+}$
clock at NIST-Boulder. Unlike most other ion clock systems $^{27}$Al$^{+}$
does not have an allowed transition accessible with current laser
technology that can be used for cooling and state detection. The NIST
group solve this problem with techniques developed for quantum information
processing with trapped ions, by co-trapping the $^{27}$Al$^{+}$
and a ``logic ion,'' either Be$^{+}$ or Mg$^{+}$. They laser cool
the logic ion, which sympathetically cools the ``clock'' ion. They
use Raman transitions to map the internal state of the clock ion to
that of the logic ion via the common motional state of the two trapped
ions, and detect the final state using the laser cooling transition
in the logic ion.

This quantum logic sectrscopy was first demonstrated in 2005\cite{QLSpectroscopy}
using the $^{1}S_{0}\rightarrow^{3}P_{1}$ transition, and used to
measure the absolute frequency of the $^{1}S_{0}\rightarrow^{3}P_{0}$
clock transition, $\nu=1\,121\,015\,393\,207\,851(6)$Hz, in 2007\cite{AlClockState}.
The initial $^{27}$Al$^{+}$ clock, using Be$^{+}$ for the logic
ion, was compared to the previously developed $^{199}$Hg$^{+}$.
They determined the frequency ratio to a remarkable 17 decimal places
by comparing the laser frequencies via a frequency comb:
\begin{equation}
\frac{\nu_{Al}}{\nu_{Hg}}=1.052871833148990438(55)\label{eq:AlHg}
\end{equation}

The optical transitions used for the clock states in both $^{27}$Al$^{+}$
and $^{199}$Hg$^{+}$ are sensitive to changes in the fine structure
constant (only weakly in Al, but more strongly in Hg), but depend
on $\mu$ and $g$ only at higher orders, which can be neglected.
Eq. \ref{eq:AlHg}, then, directly probes the change in $\alpha.$
Repeated measurements of this ratio over the course of one year give
the best laboratory limit to date on $\frac{\partial}{\partial t}\ln\alpha=\text{(\textminus}1.6\pm2.3\text{)\ensuremath{\times}}10^{-17}$yr$^{-1}$\cite{AlHgComp}.

The NIST group has since constructed a second $^{27}$Al$^{+}$clock,
replacing the Be logic ion with Mg (whose mass is closer to Al allowing
for more efficient coupling between the ions), and incorporating some
mechanical improvements. These upgrades improve the clock performance
by more than a factor of two, from an inaccuracy of $2.3\times10^{-17}$
for the Al-Be clock to $0.86\times10^{-17}$\cite{Al-MgClock}. Measurement
at this level of accuracy opens many possibilities for new physics
searches, not only through improved clock comparisons, but through
gravitational measurements, as the clock uncertainty is comparable
to the gravitational redshift for a change in elevation of $\sim10\,$cm;
the NIST group compared the frequencies of their two $^{27}$Al$^{+}$
clocks, and clearly resolved the frequency shift due to raising one
33cm above the other\cite{AlClockRelativity}. 

The other major category of optical frequency clocks use neutral atoms
in optical lattices\cite{LatticeClockReview}. Lattice clocks offer
the advantage of using many atoms rather than a single ion (using
multiple ions in a single trap can lead to large frequency shifts
due to the Coulomb interaction), but the confining lattice must be
operated at a ``magic'' wavelength where the AC Stark shifts of
the two states in the clock transition are identical. The need for
a ``magic'' lattice wavelength limits the atomic species that can
be used in a lattice clock, but lattice clocks have been demonstrated
in Sr\cite{TakamotoSrLattice,LudlowSrLattice,LeTargetSrClock}, Yb\cite{KohnoYbLattice,LemkeYbLattice},
and Hg\cite{HachisuHgLattice,PetersenHgLattice,McFerranHgClock}.

The most fully developed of these systems is strontium, with $^{87}$Sr
lattice clocks operating in Tokyo, Boulder, and Paris since 2006.
The measured transitions frequencies for the $^{1}S_{0}\rightarrow^{3}P_{0}$
clock transition, independently calibrated by comparisons to Cs primary
standards, agree to within $1.7$Hz out of $429\:228\:004\:229\,874\,$Hz.
This comparison limits both the possible change in fundamental constants
($\dot{\alpha}/\alpha=(-3.3\pm3.0)\times10^{-16}yr^{-1}$, $\dot{\mu}/\mu=(1.6\pm1.7)\times10^{-16}yr^{-1}$)
and the possible annual variation due to a coupling to gravity\cite{SrClockNetwork}.
This limit is not as tight as the Hg$^{+}$-Al$^{+}$ comparison,
\cite{AlHgComp}, but much of the uncertainty comes from the comparison
to the microwave Cs standards. Direct comparisons to other optical
frequency standards and improvements in optical clock technology should
reduce these uncertainties, as will observations over a longer period
of time.

\subsection{Other Systems}

While existing clock systems place limits on the modern rate of change
of fundamental constants, the current sensitivity will need to improve
by 1-2 orders of magnitude in order to address the question of the
spatial variation seen in quasar data. Improvements in optical clocks
may bring this sensitivity within reach, two other dramatically different
approaches might also reach the necessary sensitivity.

One method drawing on atomic clock technology is a ``nuclear clock''
using the low-energy ($\sim7.6$eV) isomer transition in $^{229}$Th,
which it may be possible to excite with VUV lasers\cite{ThProposal,ThSingleIon}.
Such a nuclear transition would offer exceptional shielding against
external perturbations, to the point where a clock using large numbers
of trapped ions\cite{ThIonCrystal}, or thorium nuclei implanted in
a solid crystal of a material such as CaF that is transparent to the
excitation light\cite{ThSolidState} might be feasible. Both trapped-ion
and solid-state proposals are estimated to reach accuracies of $10^{-19}$.
Additionally, the nuclear isomer transition might offer enhanced sensitivity
to changes in fundamental constants, though more accurate models will
be needed to determine the exact sensitivity\cite{ThSensitivityTheory}.
Several groups are working toward laser spectroscopy of the isomer
transition, though it has not yet been detected.

The other relatively new method potentially sensitive enough to detect
spatial variation in $\alpha$ uses the $4f^{10}5d6s\rightarrow4f^{9}5d^{2}6s$
transition between two nearly degenerate states of opposite parity
in dysprosium\cite{DyProposal}. This energy splitting can be measured
directly using an allowed dipole transition, and the RF transition
frequencies can be counted directly, greatly simplifying the analysis.
The energy splitting of the transition is comparable to the isotope
shift for Dy, so the frequency shift due to changing $\alpha$ has
different sign for different isotopes, removing the need for an explicit
comparison to other elements.

An initial measurement comparing the $235$MHz transition in $^{162}$Dy
with the $3.1$MHz transition in $^{163}$Dy gives a limit of $\dot{\alpha}/\alpha=(-2.7\pm2.6)\times10^{-15}yr^{-1}$\cite{CingozDy}.
The eight-month span of this measurement also places a limit on annual
variation due to a coupling to gravity\cite{FerrellDyGrav}. These
initial limits are not as good as the best clock comparison data,
but the ultimate possible sensitivity may be as high as $\dot{\alpha}/\alpha\sim10^{-18}yr^{-1}$\cite{NguyenDyProposal}.
Since those measurements, laser cooling has been demonstrated in Dy,
with both boson\cite{LuDyBEC} and fermion\cite{LuDyDFG} isotopes
cooled to quantum degeneracy, which may allow further improvements. 

The precise sensitivity to changes in $\alpha$ and other fundamental
constants depends on the details of atomic and nuclear wavefunctions.
Numerical calculations are underway in neutral atoms\cite{NeutralAlphaDot}
and highly charged ions\cite{IonAlphaDot} to determine other promising
candidates for finding new physics through precision measurement.

\section{The Value of $\alpha$}

Detecting changes in $\alpha$ is one possible way for AMO techniques
to detect new physics, but the absolute value of $\alpha$ can also
shed light on possible new physics. A recent measuremnt of the g-factor
that determines the magnetic moment due to the electron's spin ${\bf \mathbf{\boldsymbol{\mu}}}=-g\frac{e}{2m_{e}}\mathbf{S}$,
combined with a tenth-order QED calculation gives the best current
measurement of the value of $\alpha.$ A comparison of this value
and the best independent measurement of $\alpha$, based on recoil
measurements of Rb atoms, provides the most stringent test of QED
to date, and allows clear observation of the muonic and hadronic contributions
to the electron g-factor.

\subsection{The Electron g-factor}

While the techniques used are drawn from precision AMO measurements,
the key system in this case is not a naturally ocurring atom, but
an artificial one: a single electron held in a Penning trap, consisting
of an axial magnetic field and an electrostatic quadrupole field\cite{geonium}.
Precision measurement of $\alpha$ uses two sets of quantized states
due to the combination of cyclotron motion (cyclotron frequency $\nu_{c}=(e/2\pi m_{e})B_{z}$)
and the interaction between the electron spin and the trap magnetic
field. The total energy of the $n$th cyclotron state is:

\[
E_{nm_{s}}=E_{n}^{(cyc)}+E_{m_{s}}^{(spin)}=\left(n+\frac{1}{2}\right)h\nu_{c}+\frac{g}{2}h\nu_{c}m_{s}-\frac{1}{2}\hbar\delta\left(n+\frac{1}{2}+m_{s}\right)^{2}
\]
where $m_{s}=\pm\frac{1}{2}$ and $\delta$ is a relativistic correction
factor of order $10^{-9}$. 

For a Dirac point particle, $g=2$ exactly, and the $|n,m_{s}=\frac{1}{2}>$
state would be degenerate with the $|n+1,m_{s}=-\frac{1}{2}>$ state,
to within$\sim\delta$. QED corrections give $g$ for a real electron
a slightly greater value, though, leading to a small difference between
spin-up and spin-down states, expressed as an ``anomaly frequency''
$\nu_{a}$. Both cyclotron and ``anomaly'' transitions are driven
by RF fields applied to trap electrodes, and a weak coupling between
the cyclotron motion and the axial motion allows the direct measurement
of the state of the electron by picking up the axial motion with the
same electrodes.

All of the properties of this system are extremely well-controlled,
enabling the measurement of g to $0.3$ppt precision\cite{Hanneke_g-2,Hanneke_g-2_PRA}:
\begin{equation}
\frac{g}{2}=1.001\,159\,652\,180\,73(28)\label{eq:g-factor}
\end{equation}
this value, combined with a QED calculation of $g/2$ to tenth order,
involving the summing of some 13,000 Feynman diagrams, determines
the best measurement of the fine structure constant\cite{tenth_order_theory}:
\begin{equation}
\frac{1}{\alpha}=137.035\,999\,166(34)\label{eq:fine-structure-from-g}
\end{equation}
(where the uncertainty is dominated by the experimental uncertainty
in the measurement of $g$).

The same technique used to measure $g$ for the electron should work
to measure $g$ for the positron, improving upon existing tests of
CPT symmetry for leptons\cite{LeptonCPT}. A similar method has recently
been used to make a direct measurement of the magnetic moment of the
proton\cite{proton_moment_DiSciacca} (which is more technically challenging
as the moment is smaller than that of the electron by a factor of
$m_{p}/m_{e}\sim1836$), and spin flips of a single trapped proton
have been observed\cite{proton_spin_flip,proton_magnetic_details}.
These should lead to high-precision measurements of the proton (and
antiproton) $g$, allowing exceptionally precise QED and CPT tests
with hadrons.

\subsection{$\alpha$ From Atomic Recoil}

Equation \ref{eq:fine-structure-from-g} is the best measurement to
date of the fine-structure constant, but extracting $\alpha$ from
$g$ necessarily assumes the correctness of the QED calculation in
Ref. \cite{tenth_order_theory}. \emph{Testing }QED requires an independent
measure of $\alpha$ with which to calculate a theoretical value of
$g$. 

We can express $\alpha$ in terms of other well-known constants and
the mass $m_{Rb}$ of a rubidium atom: 
\[
\alpha^{2}=\frac{R_{\infty}}{2c}\frac{m_{Rb}}{m_{e}}\frac{h}{m_{Rb}}
\]
The Rydberg constant $R_{\infty}$and the mass ratio are known to
better than ppb accuracy, so an accurate measurement of the ratio
$h/m_{Rb}$ enables a direct measurement of $\alpha,$without any
assumptions regarding QED. This ratio is determined from the recoil
velocity of a Rb atom absorbing a photon of momentum $\hbar k$:

\[
v_{r}=\frac{\hbar k}{m_{Rb}}
\]
The recoil velocity has been measured to ppb accuracy\cite{RbRecoil_Bouchendira}
using a Ramsey-Bordé interferometer\cite{Ramsey-Borde}, a variant
of the Ramsey interferometry method described in Section \ref{sub:Ramsey-Interferometry},
combined with the Bloch oscillation technique for transferring large
amounts of momentum to a sample of atoms. 

A sample of ultracold Rb atoms in the $F=2$ hyperfine ground state
are subjected to a $\pi/2$ pulse of a Raman transition that prepares
a coherent superposition of $F=2$ and $F=1$ hyperfine levels, where
the atoms moved to the $F=1$ state have also acquired a velocity
of two photon recoils. Some time later, a second $\pi/2$ pulse is
applied, which leads to Ramsey fringes within the velocity distribution
of the $F=1$ atoms. A second pair of $\pi/2$ pulses some time later
returns the atoms to the $F=2$ state, and completes the Ramsey-Bordé
interferometer, converting the fringes in the velocity distribution
to fringes in the transition probability, as a function of the frequency
of the final $\pi/2$pulses.

In the absence of any other interactions, such an interferometer is
sensitive to accelerations or rotations of the atoms during the time
between the sets of $\pi/2$ pulses. To measure the recoil velocity,
they add an acceleration stage between the $\pi/2$ pulses, illuminating
the atoms with a pair of counter-propagating lasers with a frequency
offset between them. This can be viewed either as trapping the atoms
in an optical lattice that is accelerated as the frequency offset
is swept, or as performing a series of $N$ Raman transitions starting
and ending in the same internal state, but increasing the velocity
by $2v_{r}$ for each transition. This process is analogous to the
Bloch oscillations of an electron in a solid subjected to an electric
field. 

At the end of the acceleration phase, the atoms have acquired a velocity
of $2Nv_{r}$, leading to a Doppler shift of the final Ramsey-Bordé
inteference pattern of $\Delta\omega=2Nkv_{r}$. Comparing the shift
for upward- and downward-accelerated atoms removes the effect of gravity,
determining $\alpha$ to $0.66$ppb\cite{RbRecoil_Bouchendira,tenth_order_theory}:

\[
\frac{1}{\alpha}=137.035999037(91\text{)}
\]
Using this value for $\alpha$ to calculate $g$ gives:
\[
\frac{g}{2}=1.001\,159\,652\,181\,82(78)
\]

These values agree Eq. \ref{eq:fine-structure-from-g} and \ref{eq:g-factor},
confirming the QED calculation at the ppb level. This accuracy is
sufficient to show the contributions of muonic and hadronic terms
to the value of $g$ (assuming the correctness of the elecron-only
parts of QED): a calculation omitting these terms differs from the
result of Eq. \ref{eq:fine-structure-from-g} by roughly $2.5\sigma$\cite{RbRecoil_Bouchendira}.

\section{Electron EDM}

The final class of AMO new physics searches are those seeking a permanent
electric dipole moment (EDM) of the electron. Purcell and Ramsey\cite{Purcell_Ramsey}
first noted in the 1950's that the existence of a permanent EDM for
a fundamental particle would violate time-reversal symmetry: the spin
magnetic moment of a particle is odd under time reversal, where an
EDM would be even. The existence of a T-violating EDM implies CP-violation
elsewhere, assuming CPT symmetry. Known sources of CP-violation within
the Standard Model predict EDMs far too small to measure exerimentally,
in particular, an electron EDM with a magnitude of $d_{e}\sim10^{-40}e-cm$.
Most theories of physics beyond the Standard Model introduce new sources
of CP-violation, though, and predict an electron EDM many orders of
magnitude larger: $d_{e}\sim10^{-25}-10^{-30}e-cm$\cite{old_edm_theory,new_edm_theory}.

The electron EDM cannot be detected with a free electron (which merely
moves in response to an electric field), but can lead to small energy
shifts in the states of atoms or molecules in an electric field. In
the simplest approximation, one would expect electrons in an atom
to shift so as to cancel an applied electric field, but relativistic
effects prevent a perfect cancellation, and for some heavy atoms can
even produce an enhancement of the electric field experienced by an
electron within an atom. The best atom-based measurement used thallium,
where the applied field is enhanced by a factor of $\sim580$\cite{edm_thallium_Regan}.
More recent EDM searches use polar molecules, where the internal electric
field of the molecule can be orders of magnitude larger than the enhanced
field within an atom. The effective field for a recent measurement
using YbF\cite{edm_YbF_Hudson} molecules was 220 times larger than
the field used in the Tl measurement, for a much smaller applied field
in the lab.

\begin{figure}
\includegraphics[scale=0.4]{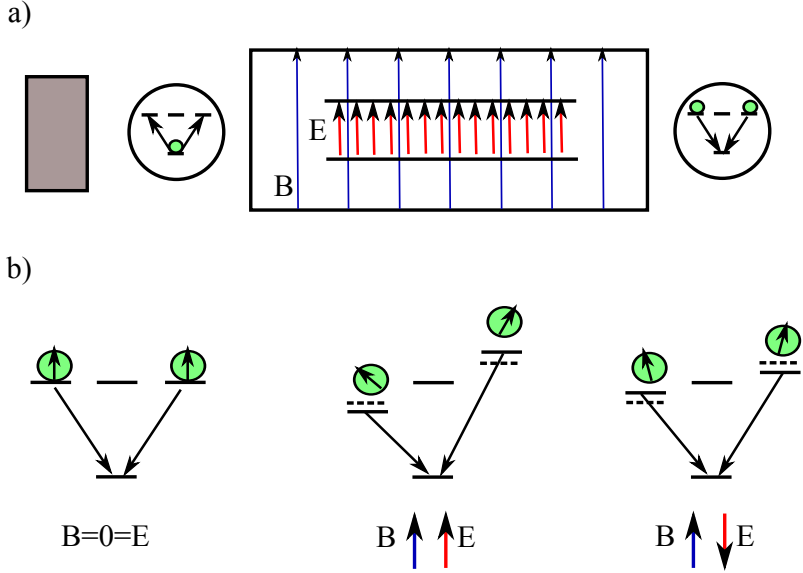}

\caption{a) Experimental schematic of the electron EDM search. Atoms or molecules
are prepared in an initial superposition state, then allowed to evolve
freely in a region of uniform electric and magnetic fields. After
the free evolution, the probability of returning to the initial state
depends on the relative phases of the two parts of the superposition,
giving rise to Ramsey fringes. b) Schematic of the interference measurement.
In the absence of an applied field (left), the two states in the superposition
are degenerate, and remain in phase. With parallel E and B fields
(center) the Zeeman shift (dashed line) plus the EDM shift give a
phase difference between the two components. Reversing the direction
of the E-field (right) does not change the Zeeman shift, but reverses
the direction of the EDM shift, changing the phase difference. The
resulting change in Ramsey fringes on reversal of the E-field }

\end{figure}

In both atomic and molecular experiments, the EDM measurement uses
a variant of Ramsey interferometry. Rather than preparing a superposition
of two different energy states, a beam of atoms or molecules are excited
by a ``$\pi/2$'' pulse to a coherent superposition of two Zeeman
sublevels of the same hyperfine state ($F=1\; m_{f}=0\rightarrow F=1,\; m_{f}=\pm1$
in Tl, and $F=0\; m_{f}=0\rightarrow F=1,\; m_{f}=\pm1$ in YbF).
A uniform magnetic field applied during the free evolution stage creates
a phase difference between the two sublevels due to their different
Zeeman shifts. This phase difference leads to interference between
the two populations when the second ``$\pi/2$'' pulse attempts
to return the population to the initial $m_{f}=0$ state, producing
Ramsey fringes in the transition probability that are detected with
a final state measurement.

To look for an EDM, an electric field is applied either parallel or
anti-parallel to the magnetic field. A non-zero electron EDM produces
a shift in the energy levels, and thus the interference fringes, that
depends on the direction of the E-field. A complete set of measurements
consists of many repetitions of the experiment with field directions
and magnitudes switched (cycling through 256 combinations of experimental
parameters in Tl, and 512 in YbF). Measurements in $^{205}$Tl limit
$|d_{e}|\leq1.6\times10^{-27}e-cm$\cite{edm_thallium_Regan}, and
YbF provides the best current limit, $|d_{e}|\leq1.05\times10^{-27}e-cm$\cite{edm_YbF_Hudson}.

These measurements already place stringent constraints on theories
of physics beyond the Standard Model, ruling out some of the simpler
extensions. A sensitivity improvement of another 1-2 orders of magnitude
should either definitively detect an electron edm or severely restrict
the most popular Standard Model extensions. Such an improvement may
be possible in the YbF experiment, for example by cooling the molecular
beam source\cite{cold_YbF}, which will dramatically improve the measurement
statistics. Another possibility is to move to a different molecule
offering better sensitivity; there are numerous such proposals, the
most fully developed of which uses thorium monoxide\cite{ThO_ACME}.
As of June 2012, the ACME collaboration reports that they are taking
preliminary EDM data, and hope improve the YbF limit by an order of
magnitude in the near future\cite{ACME_DAMOP}. Detection of a non-zero
electron EDM would imply the existence of new physics at the TeV level,
detected using low-energy lasers and modest electric and magnetic
fields.

\section{Conclusion}

The experiments described here represent a small subset of all precision
AMO searches for new physics. Numerous other experiments are underway,
testing fundamental symmetries with atomic clocks, searching for spin-dependent
effects with atomic magnetometers, using atom interferometry to probe
gravitational physics, and many others. While it is not possible to
describe all of these experiments, the current selection should provide
some sense of the poential and the power of precision-based searches
for physics beyond the Standard Model.

\section*{Acknowledgements}

Thanks to Tom Swanson and Dave Phillips for helpful comments on a
draft of this article. This article grew out of an invited talk at
the 2011 DAMOP meeting.

\end{document}